\documentclass[12pt,preprint]{aastex}

\begin{document}

\title{Using Star Spots to Measure the Spin-orbit Alignment of Transiting Planets}

\author{	Philip A. Nutzman,
		Daniel C. Fabrycky,
		Jonathan J. Fortney\altaffilmark{1}
		}

\altaffiltext{1}{Department of Astronomy and Astrophysics, University of California, Santa Cruz}

\email{pnutzman@ucolick.org}

\begin{abstract}

Spectroscopic follow-up of dozens of transiting planets has revealed the degree of alignment between the equators of stars and the orbits of the planets they host.  Here we determine a method, applicable to spotted stars, that can reveal the same information from the photometric discovery data, with no need for follow-up.   A spot model fit to the global light curve, parametrized by the spin orientation of the star, predicts when the planet will transit the spots. Observing several spot crossings during different transits then leads to constraints on the spin-orbit alignment.   In cases where stellar spots are small, the stellar inclination, $i_s$, and hence the true alignment, rather than just the sky projection, can be obtained.  This method has become possible with the advent of space telescopes such as \emph{CoRoT} and \emph{Kepler}, which photometrically monitor transiting planets over a nearly continuous, long time baseline.  We apply our method to CoRoT-2, and find the projected spin-orbit alignment angle, $\lambda= 4.7^\circ \pm 12.3^\circ$, in excellent agreement with a previous determination that employed the Rossiter-McLaughlin effect.  The large spots of the parent star, CoRoT-2, limit our precision on $i_s$: $84^\circ \pm 36^\circ$, where $i_s < 90^\circ ~(> 90^\circ)$ indicates the rotation axis is tilted towards (away from) the line of sight. 

\end{abstract}

\section{Introduction}

Transit observations of several systems, e.g., HD 209458 (Silva 2003), TrES-1 (Charbonneau et al. 2007), HD 189733 (Pont et al. 2007), and CoRoT-2 (Alonso et al. 2008, Silva-Valio et al. 2010), reveal anomalous flux rises during transit, most likely the result of the planet occulting a dark spot on the stellar surface.   Spots on the stellar surface complicate the estimation of planetary parameters from transit photometry and can lead to errors in planet size measurements if not accounted for (e.g., Czesla et al. 2009).  

On the other hand, spots introduce structure into out-of-transit and in-transit observations, which may offer new opportunities to learn about the planet, its orbit, and the host star.  Silva-Valio (2008) pointed out the possibility of estimating the rotation period of transit host-stars, if an observer is able to catch the transiting planet occulting the same spot on two consecutive transits.  The key to the method is measuring the displacement of the “spot perturbation” along the transit chord and modeling the displacement as due to rotation of the star.  Applying the method to Hubble Space Telescope observations of HD 209458b,  and implicitly assuming alignment in the projected spin-orbit alignment angle ($\lambda=0^\circ$) and a stellar inclination of $i_s=90^{\circ}$,  Silva-Valio (2008) constrained the rotation period of the star to be either 9.9 or 11.4 days.  Rotation period estimates from long-term photometric observations of out-of-transit variability in HD 209458 find a value consistent with the 11.4 day estimate.  Dittmann et al. (2009), in a similar analysis of TrES-1b, generalized the method by considering when $\lambda \neq 0^{\circ}$ and $i_s \neq 90^{\circ}$, but found that this freedom introduced large uncertainty in the estimation of the rotation period.  

We propose to reverse the chain of inference.  Because space-based transit monitoring missions, such as \emph{Kepler} and \emph{CoRoT}, observe each target over a nearly continuous, long time-baseline, we can robustly determine the rotation period and rotational phase of the occulted spot at the time of each transit by fitting a spot model to the “global” light curve.  By comparing the position of the spot perturbation along the transit chord with the spot's rotational phase, we can constrain $\lambda$ and $i_s$.  

Constraints on the projected spin-orbit alignment angle via Rossiter-McLaughlin (RM) measurements (e.g., Gaudi \& Winn 2007) have yielded unique science.  For instance, it appears that misaligned systems are preferentially found around F-type stars (Schlaufman 2010) which may point to a connection between alignment and the larger convective atmospheres of cooler stars (Winn et al. 2010).  Meanwhile, attempts are being made to understand the relative importance of two (or more) modes of planetary migration, which bring planets from their formation zone into close-in orbits (Morton \& Johnson 2010).  Planet-planet scattering followed by tidal circularization (Chatterjee et al. 2008) and Kozai oscillations followed by tidal circularization (Fabrycky \& Tremaine 2007), both lead to misalignment, while migration through a gaseous disk should, as currently understood (Lin et al. 1996), lead to alignment.

In this paper, we describe our method to constrain the spin-orbit alignment and demonstrate an application to the CoRoT-2 system.  A similar concept has been developed by Sanchis-Ojeda et al. (2011) and was applied to WASP-4b, finding $\lambda= -1^{+14}_{-12}$ degrees.  Our method and analysis was developed independently of Sanchis-Ojeda et al. (2011), whose analysis was reported while this manuscript was in preparation. In \S 2, we discuss the \emph{CoRoT} observations of CoRoT-2, measure the timing of in-transit spot-crossings, and estimate the timing uncertainty. In \S 3, we describe a spot model to the global CoRoT-2 photometry, and our model for the in-transit spot timings.  We confront the spot timing model with the CoRoT-2 spot timing measurements. We conclude in \S 4 with a discussion of our results and contrast our method with that of Sanchis-Ojeda et al. (2011).  

\section{Data}

\subsection{Observations}
CoRoT-2 was continuously monitored over approximately 140 days by the Convection, Rotation, and Planetary Transits (\emph{CoRoT}) space telescope (Alonso et al. 2008).  The first week of observations was conducted with a sampling of 512 s, and 32 s for the remainder of the observations.  We consider only the data at 32 s sampling. For this data set, 78 transits were observed.  We identified the transits using the ephemeris of Alonso et al. (2008), $T_E= E \times 1.7429964 + 
2454237.53562$ BJD.  We refer to transits by the transit number $E$ as defined by the ephemeris above. For the purposes of fitting a global spot model to the data, we removed in-transit observations and re-binned the data to 512 s resolution.  For analysis of transit data, we retain the 32 s sampling.  The relative root-mean-square (rms) photometric scatter at 32 s sampling is 720 parts-per-million (ppm).

\subsection{Spot-crossing time measurements}

For each transit, we select an 8-hour window bracketing the time of mid-transit.  The transit duration is 1.94 hours (defined here as the time between the crossings of the planet center over the stellar limb).  We derive a linear fit using the out-of-transit data within the 8-hour time window, and normalize the data (including the in-transit observations) by dividing off the linear fit.  We then determine the residuals to the transit model, using transit parameters derived by Gillon et al. (2010).  Because of the difficulty of identifying spot crossings that occur on the limb of the star, we only attempt timing measurements for spot-crossings which show a clear signal between transit ingress and egress.

The shape of a spot perturbation to a transit light curve depends on the relative sizes of the planet and star spot, and the brightness distribution of the star spot.  
The ability to resolve the time of spot crossing is limited by the planet size and spot size, with the best case occurring when the spot is much smaller than the planet.  In the case of CoRoT-2, the spot (or perhaps, complex of small spots) in consideration is at least as large as the planet.  This is evident because the durations of the spot perturbations are longer than transit ingress/egress.  For simplicity, we identify the spot crossing timing as the time of maximum spot perturbation.  In the modeling of \S 3, we associate this time with the time of minimum projected separation between planet and spot centers.  This correspondence is of course a simplifying approximation, which may fail due to limb-darkening and foreshortening of the stellar surface, for irregularly shaped spots, and for spots on the limb of the star.  We mitigate these effects by only considering spot perturbations in the flat-bottomed portion of the transit light curve.  To compensate for the diminution of the spot-crossing signal caused by limb-darkening and foreshortening, we normalize the residuals to the transit model, dividing by the ``instantaneous'' transit depth $(1 - F_{transit})$, where $F_{transit}$ is the model specified by the parameters of Gillon et al. (2010).  Even after these steps, the approximation is imperfect, but we expect any errors introduced to be small compared to the timing uncertainty.

To measure the timing and timing uncertainty of the spot-crossing, we have used the following non-parametric procedure: 1.) We construct 10,000 realizations of the observed data set by adding to each observed value a random number drawn from a normal distribution with mean 0 and standard deviation equal to the rms photometric scatter of 720 ppm.  2.) We normalize the residuals, dividing by the ``instantaneous'' transit depth $(1 - F_{transit})$.  This normalization is performed on each of the 10,000 realizations of the data set.  3.)  For each normalized realization, we note the time of maximum residual.  4.) We take the spot-crossing timing and uncertainty to be the mean and standard deviation of the 10,000 noted times.  

\section{Method with application to CoRoT-2} 

\subsection{Global light curve spot model}

Characterizing stellar spots by fitting a spot model to the global light curve is complicated by a large number of free parameters, degeneracies between parameters, spot evolution, and the need to make assumptions about spot shape.  Our goal with spot modeling is the robust determination of the rotational phase of each spot at any given moment of time, which is possible despite the above concerns.  Because star spots may change shape, migrate, appear, and disappear over the entire 135 day observing window of CoRoT-2, we analyze a shorter 75-day segment (from BJD 2454247.5 to 2454322.5 of the CoRoT-2 data set), which we further subdivide into three 25-day subsets.  We determine an independent best-fit spot model for each of the three subsets, thus allowing for spot evolution from one data subset to the next.  The 25-day window length is a compromise between the need for several stellar rotations in order to reliably determine spot longitudes, and the need to model over a time window that is ideally no longer than the evolutionary timescale of the starspots.  We note that Lanza et al. (2009) find a typical spot lifetime of approximately 55 days for CoRoT-2, and a cyclic oscillation in the total spotted area with a period of approximately 29 days.  

We adopt the analytical spot model of Dorren (1987), which assumes circular spots and linear limb-darkening.  Following Fr{\"o}hlich et al. (2009), we adopt a 3-spot model for CoRoT-2.  We allow for differential rotation by fitting the period of each spot independently.  Our model includes five parameters for each spot: spot radius, latitude, CoRoT bandpass spot-star brightness ratio, and the period and epoch of rotation.  The model also includes a linear limb-darkening parameter and the stellar spin inclination.  One possible improvement to the spot modeling would be to extend the Dorren (1987) analytical model to include quadratic limb-darkening, however, we opted for the relative simplicity of linear limb-darkening. We expect that the differences in the estimated rotational phases resulting from an improved model would have insignificant effect on the determination of the spin-orbit alignment, given the relatively large uncertainties in the spot-crossing timings.

We searched for the best-fit solution to each 25-day subset using a downhill simplex method.  The combined, 75-day best-fit model is depicted in Figure 1.  Previous attempts to fit a single model to the entire 75-day time series led to a significantly worse fit to the global light curve than that of the three-subset model adopted in this study.  While the fit is imperfect, the local minima and maxima are reproduced accurately, indicating a reliable determination of the spot rotational phases.   
The best-fit model (in each of the three data subsets) is characterized by one spot with period near 4.94 d, and two spots with slightly different periods near 4.54 d, which are separated in rotational phase by approximately $155^{\circ}$ at BJD BJD 2454247.5, though drifting apart due to differing rotation periods.   The latter two spots correspond well with the two active longitudes described in Lanza et al. (2009).

Upon finding the best-fitting spot model, we determine the rotational phase, $\alpha$, of each spot at the time of each transit.  We define a spot's $\alpha$ so that it ranges from $-180^\circ$ to $180^\circ$ and equals zero when the spot is along the central meridian coincident with the projected stellar spin axis.  In this paper, we will focus on one of the three fitted spots, which shows evidence of being occulted by the planet in several transits.  Over the observational window that we analyze, we note that during each transit for which the rotational phase of this spot was between $-70^\circ$ and $70^\circ$, there is a clear spot-crossing bump in the transit light curve.  Arranging the transit light curves in order of increasing $\alpha$ for this spot, the spot-crossing bumps recur later and later along the transit light curve (see Figure 2).  The simplest explanation for this phenomenon is that the planet's orbit is prograde and relatively well aligned with the projected stellar spin axis, and that the bumps shown in Figure 2 represent the same spot being occulted by the planet multiple times.  In our modeling below, we assume that each of the perturbations in the transit light curves is due to this spot.  In Table 1 we list the rotational phases of this spot at the time of transit center, for each of the transits that we analyze below.    

\subsection{Spot Crossing Model}

Given the rotational phase and latitude, $l$, of a spot, the stellar spin inclination, $i_s$, the projected spin-orbit alignment angle, $\lambda$, and transit impact parameter, $b = a/R_{\star} \cos{i_p}$, we can predict the time during transit of maximum spot perturbation.  In an analysis of CoRoT-2 transit photometry, Gillon et al. (2010) find $b=0.221^{+0.017}_{-0.019}$.  The uncertainty in $b$ is insignificant for our purposes, so we fix $b=0.221$.  We assume that the time of maximum spot perturbation occurs when the transiting planet is at minimum sky-projected separation from spot center (see \S 2.2). 

Any given spot-crossing model is specified by three parameters, $\lambda$, $i_s$, and $l$.  We do not enforce consistency of $i_s$ and $l$ with best-fit value determined from the spot model fit to the global light curve, as these values are poorly constrained by the global photometry. Our $\chi^2$ statistic is defined as follows:
\begin{equation}
\chi^2 = \sum_{E=1}^{8} \frac{ (T_E(\lambda,i_s,l) -t_E)^2}{\sigma_E^2} + P_E,
\end{equation}
where $T_E(\lambda,i_s,l)$ is the model predicted spot-crossing timing for transit $E$, $t_E$ is the measured timing, and $\sigma_E$ is the measurement uncertainty (see Table 1).  $P_E$ is a penalty term which is zero whenever the spot center at the time of transit is within $30^\circ$ spherical distance of the transit chord, and a punitively large number (we picked 99) otherwise.  Because we do not know exactly how large the spot is, we chose a conservatively large value of $30^\circ$ for the required spot proximity.  

 The intuition behind the goodness-of-fit of a spot-crossing model is demonstrated in Figure 3. For a good fit, a model must accurately predict both the existence and timings of observed spot crossings.  In the misaligned example of Figure 3, the starspot on the left would not produce a perturbation, thus incurring a penalty term if one was indeed observed; whereas in the aligned example, the starspot on the left would produce a maximum perturbation just after ingress.  In this way we can distinguish misalignments of 10s of degrees.

We emphasize that, given the large and uncertain star spot sizes on CoRoT-2, the latitude of spot center is not constrained to be in the band occulted by the disk of the planet.  This limitation, combined with the large spot-crossing timing uncertainties, leads to a degeneracy of $i_s$ with $l$; almost any $i_s$ can be allowed by adjusting the spot latitude, such that the spot is near the occulted band.  Note that in the limit of a spot that is much smaller than the planet, we would require the projected distance between spot center and the transit chord to be less than the radius of the planet.  This constraint, especially if accompanied by well determined spot-crossing timings, would allow for the degeneracy between $i_s$ and $l$ to be broken, and allow for a reliable determination of the stellar spin inclination.

To determine the posterior probability distribution for the parameters, we employed a Markov Chain Monte Carlo (MCMC) method, with acceptance probability calculated with the likelihood $\exp(-\chi^2/2)$.  We assumed prior distributions that are uniform in $\lambda$, $\cos(i_s)$, and $\sin(l)$.   We take the measured value and one-sigma uncertainty for each parameter to be the mean and standard deviation of the MCMC samples.  We find $\lambda = 4.7^\circ \pm 12.3^{\circ}$.  $i_s$ and $l$ are very poorly constrained and the MCMC samples demonstrate strong degeneracy between these two parameters.  Nevertheless, we report these values for completeness: $i_s = 84^\circ \pm 36^{\circ}$ and $l = 14^\circ \pm 30^{\circ}$, where $i_s < 90^\circ$ indicates the rotation axis tilted towards the line of sight and $i_s > 90^\circ$ indicates the rotation axis tilted away from the line of sight.   

To see how important the influence of the penalty term is on the determination of $\lambda$, we repeated the analysis using the $\chi^2$ statistic without the penalty term.  We find $\lambda = 0.3^\circ \pm 16.8 ^{\circ}$.  As evident from the larger error bar, the penalty term is an important, though not dominant constraint within our method.

\section{Discussion}

In this paper, we have described a new method to determine the spin-orbit alignment of a transiting planet, and applied it to CoRoT-2b, finding $\lambda = 4.7 \pm 12.3^{\circ}$.  This planet was also the subject of Rossiter-McLaughlin measurements by Bouchy et al. (2008), who found $\lambda = 7.2 \pm 4.5^{\circ}$.  Compared to the results of Bouchy et al. (2008), our result is consistent, though less precise.  We also placed a weak constraint on the stellar inclination, $i_s = 84 \pm 36^{\circ}$.  Although this constraint is uninformative, we expect that applying our method to systems with starspots that are smaller than the planet will yield accurate determinations of the stellar inclination.  This possibility illustrates one advantage of our method over the RM method, which is entirely insensitive to differences in the stellar inclination.  

Our method possesses other advantages over RM measurements.  RM measurements require costly radial velocity follow-up, while our method makes use of ``free'' data, in the sense that the data was already obtained for other purposes.  RM measurements are generally only possible for stars with relatively large projected rotational velocities, ruling out slow rotators, or relatively fast rotators that are nearly pole-on.  The RM method requires relatively large transit depths; the smallest planet that has succumbed to Rossiter-McLaughlin measurements is the Neptune-sized HAT-P-11b (Winn et al. 2010).  While our method also favors large transit depths, we expect that the photometric capability and data provided by the \emph{Kepler} mission should allow for constraints on the spin-orbit angle of systems with planets smaller than HAT-P-11b.

The method described in this paper is similar, in concept, to that described by Sanchis-Ojeda  et al. (2011), and applies similar geometrical constraints.  In contrast to Sanchis-Ojeda et al. (2011), our method relies on global light curve measurements of the rotational period and phase of the occulted spot.  This has the disadvantage of requiring a continuous, long time-baseline of observations, which is generally only possible with space-based photometric monitoring, such as provided by the \emph{CoRoT} and \emph{Kepler} missions.  However, there is a great advantage to knowing where the spot is on the stellar surface.  For instance, only one transit of a planet over a spot is required to tell rough spin-orbit alignment.  If we know, from out-of-transit monitoring, that the spot is on the approaching side of the star (i.e., negative rotational phase), then a spot-crossing near the beginning of the transit indicates a probable prograde orbit, while a spot-crossing near the end of the transit indicates a probable retrograde orbit (with some exceptions for nearly pole-on stellar rotation).  To achieve good precision, of course, our method favors multiple spot-crossings at varying spot phases.

We gratefully acknowledge the helpful suggestions of our referee, A. Lanza.

\bibliographystyle{apj}

\begin{figure}
  \includegraphics[width=5.5in]{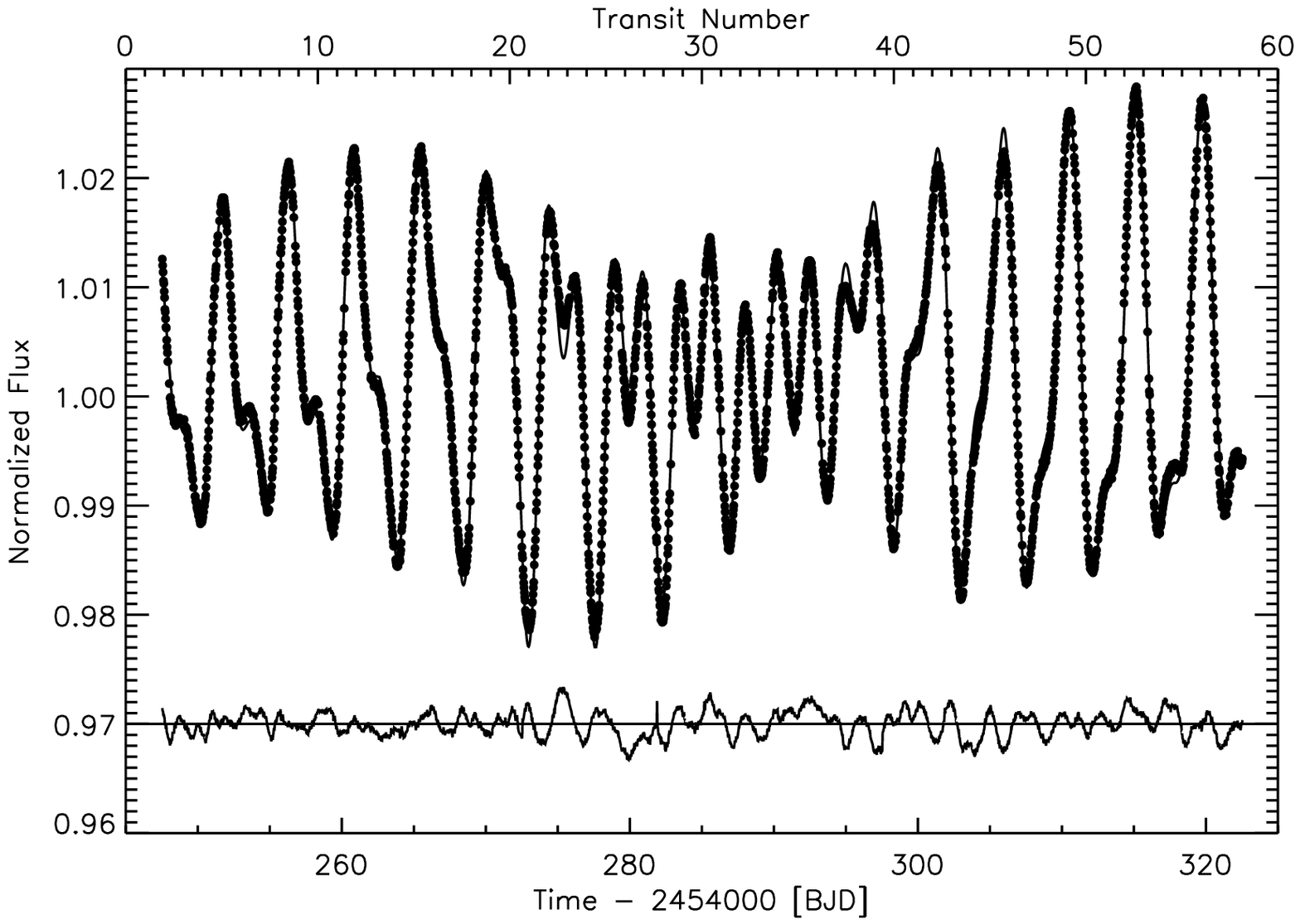}
  \caption%
  {Global light curve of CoRoT-2 over the time window we have analyzed.  The solid curve depicts the best-fit spot model.  The transits have been cut out of this data.  Below are residuals to the best-fit model, with a constant offset to +0.97.}
\end{figure}

\begin{figure}
  \includegraphics[width=5.5in]{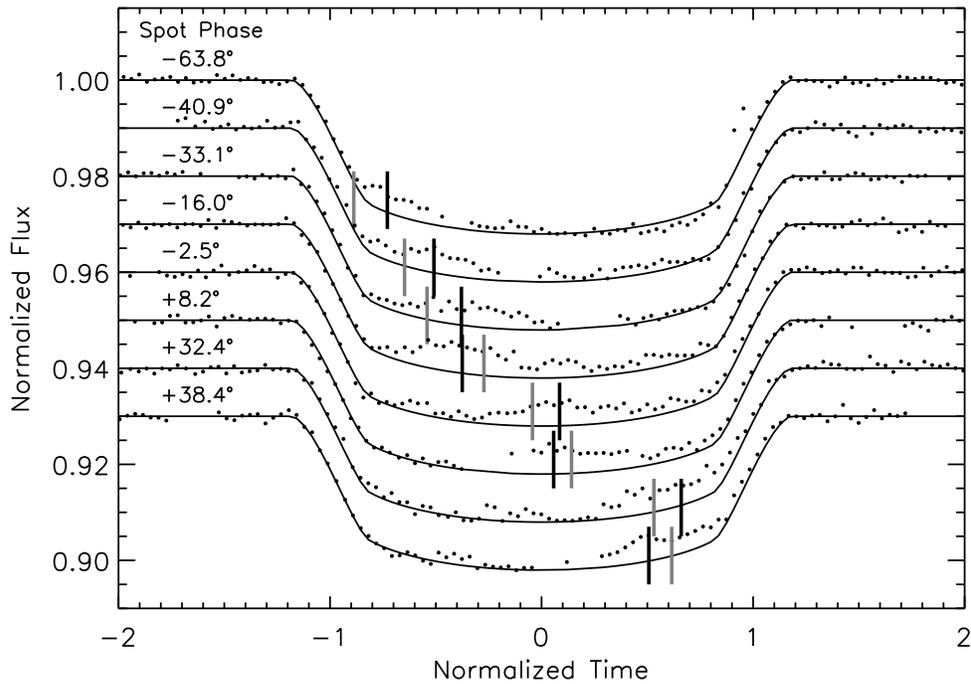}
  \caption
  {A series of transit light curves shown in order of increasing spot rotational phase.  Generally, the spot is occulted at progressively later positions in the transit light curve, with black bars indicating the spot-crossing timing as measured with the technique described in \S 2.2.  The gray bars indicate the expected spot-crossing time for a well aligned planet with, $\lambda=0$, $i_s= 90^\circ$.  The time axis is scaled so that ingress and egress, defined here by the time when the planet center crosses the stellar limb, occur at -1 and 1.}
\end{figure}

\begin{figure}
  \includegraphics[width=5.5in]{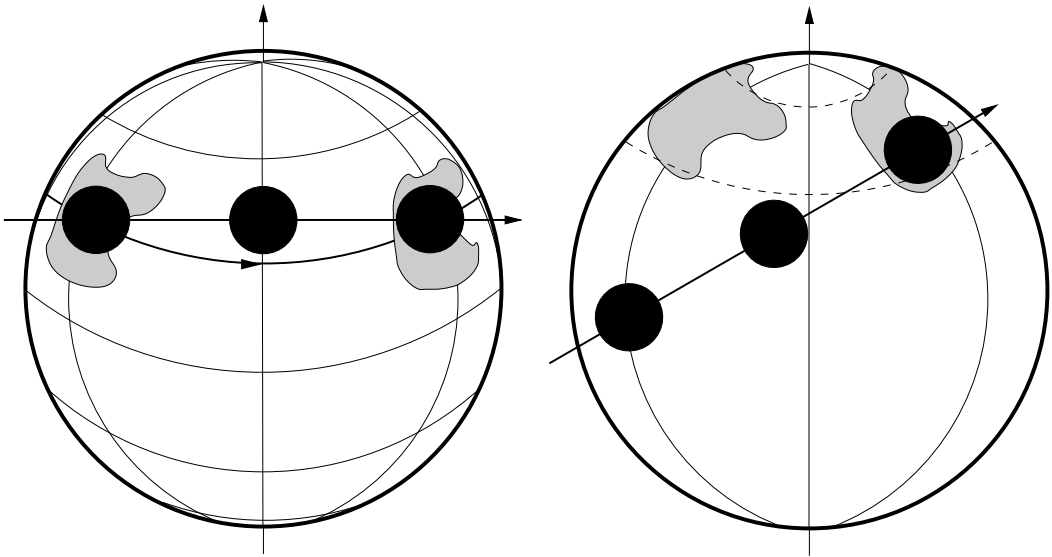}
  \caption
  {Schematic of spot crossings for a well-aligned (left) and misaligned (right) planetary orbit, with spot (gray) shown before and after a $120^\circ$ rotation of the star.  The relative size and impact parameter of the planet (black) reflect that of the CoRoT-2 system.  A planet must be well-aligned to exhibit spot crossings along the entire transit chord.  The exact timing of a spot crossing during transit, given the rotational phase of the spot, further constrains $\lambda$, while $i_s$ is largely degenerate with the spot latitude. }
\end{figure}

\begin{deluxetable}{lcc}
\tabletypesize{\scriptsize}
\tablecaption{Spot Rotational Phases, Spot-crossing Timings and Uncertainties}
\tablewidth{0pt}

\tablehead{
\colhead{Transit Number, E\tablenotemark{a}} & \colhead{Spot Rotational Phase [$^\circ$]} & \colhead{Spot-Crossing Timing\tablenotemark{b}}
}

\startdata
 $37$   & -63.8	&  $-0.729	\pm 0.080 $	\\
 $45$   & -40.9	&  $-0.509 	\pm 0.074 $	\\
 $32$   & -33.1	&  $-0.379 	\pm 0.108 $	\\
 $19$	& -16.0	&  $-0.375 	\pm 0.115 $	\\
 $27$    & -2.5	&  $0.085 	\pm 0.090 $	\\
 $14$   & 8.2	&  $0.057 	\pm 0.064 $	\\
 $9$	& 32.4	&  $0.660 	\pm 0.156$	\\	
 $22$	& 38.4	&  $0.507 	\pm 0.079$	\\
\enddata
\tablenotetext{a}{Transit number, E, correspond to transit times given by the ephemeris of Alonso et al. (2008): $T_E= E \times 1.7429964 + 2454237.53562$ d }
\tablenotetext{b}{The timings have been scaled such that ingress and egress, defined here as the time the planet center crosses the stellar limb, correspond to -1 and 1.  The duration of transit, from scaled time -1 to 1, is 1.94 hours.}
\end{deluxetable}

\end{document}